\documentstyle[epsf]{mn}

\newif\ifAMStwofonts

\newcommand{\hi}{H\,{\sc i}}
\newcommand{\hei}{He\,{\sc i}}
\newcommand{\msol}{\mbox{${\rm M}_\odot$}}
\newcommand{\fb}{\mbox{$f_{\rm B}$}}
\newcommand{\hubble}{\mbox{$\rm km\, s^{-1}\, Mpc^{-1}$}}
\newcommand{\kms}{\mbox{$\rm km\, s^{-1}$}}
\newcommand{\mhi}{\mbox{$M_{\rm HI}$}}
\newcommand{\icmsq}{\mbox{$\rm cm^{-2}$}}

\ifoldfss
  \ifCUPmtlplainloaded \else
    \NewTextAlphabet{textbfit} {cmbxti10} {}
    \NewTextAlphabet{textbfss} {cmssbx10} {}
    \NewMathAlphabet{mathbfit} {cmbxti10} {} 
    \NewMathAlphabet{mathbfss} {cmssbx10} {} 
  \fi
  \ifAMStwofonts
    \ifCUPmtlplainloaded \else
      \NewSymbolFont{upmath} {eurm10}
      \NewSymbolFont{AMSa} {msam10}
      \NewMathSymbol{\upi}     {0}{upmath}{19}
      \NewMathSymbol{\umu}     {0}{upmath}{16}
      \NewMathSymbol{\upartial}{0}{upmath}{40}
      \NewMathSymbol{\leqslant}{3}{AMSa}{36}
      \NewMathSymbol{\geqslant}{3}{AMSa}{3E}

    \fi
  \fi
\fi 

\ifnfssone
  \newmathalphabet{\mathit}
  \addtoversion{normal}{\mathit}{cmr}{m}{it}
  \addtoversion{bold}{\mathit}{cmr}{bx}{it}
  \newmathalphabet{\mathbfit} 
  \addtoversion{normal}{\mathbfit}{cmr}{bx}{it}
  \addtoversion{bold}{\mathbfit}{cmr}{bx}{it}
  \newmathalphabet{\mathbfss} 
  \addtoversion{normal}{\mathbfss}{cmss}{bx}{n}
  \addtoversion{bold}{\mathbfss}{cmss}{bx}{n}
  \ifAMStwofonts
    \ifCUPmtlplainloaded \else
      \UseAMStwoboldmath
      \makeatletter
      \new@mathgroup\upmath@group
      \define@mathgroup\mv@normal\upmath@group{eur}{m}{n}
      \define@mathgroup\mv@bold\upmath@group{eur}{b}{n}
      \edef\UPM{\hexnumber\upmath@group}
      \new@mathgroup\amsa@group
      \define@mathgroup\mv@normal\amsa@group{msa}{m}{n}
      \define@mathgroup\mv@bold\amsa@group{msa}{m}{n}
      \edef\AMSa{\hexnumber\amsa@group}
      \makeatother
      \mathchardef\upi="0\UPM19
      \mathchardef\umu="0\UPM16
      \mathchardef\upartial="0\UPM40
      \mathchardef\leqslant="3\AMSa36
      \mathchardef\geqslant="3\AMSa3E
    \fi
  \fi
\fi 

\ifnfsstwo
  \DeclareMathAlphabet{\mathbfit}{OT1}{cmr}{bx}{it}
  \SetMathAlphabet\mathbfit{bold}{OT1}{cmr}{bx}{it}
  \DeclareMathAlphabet{\mathbfss}{OT1}{cmss}{bx}{n}
  \SetMathAlphabet\mathbfss{bold}{OT1}{cmss}{bx}{n}
  \ifAMStwofonts
    \ifCUPmtlplainloaded \else
      \DeclareSymbolFont{UPM}{U}{eur}{m}{n}
      \SetSymbolFont{UPM}{bold}{U}{eur}{b}{n}
      \DeclareSymbolFont{AMSa}{U}{msa}{m}{n}
      \DeclareMathSymbol{\upi}{0}{UPM}{"19}
      \DeclareMathSymbol{\umu}{0}{UPM}{"16}
      \DeclareMathSymbol{\upartial}{0}{UPM}{"40}
      \DeclareMathSymbol{\leqslant}{3}{AMSa}{"36}
      \DeclareMathSymbol{\geqslant}{3}{AMSa}{"3E}
    \fi
  \fi
\fi 

\ifCUPmtlplainloaded \else
  \ifAMStwofonts \else 
    \def\upi{\pi}
    \def\umu{\mu}
    \def\upartial{\partial}
  \fi
\fi

\title[A Targeted Survey for H\,{\normalsize\it I} Clouds in Galaxy Groups]
       {A Targeted Survey for H\,{\Large\bf I} Clouds in Galaxy Groups}
\author[Martin A. Zwaan]
       {Martin A. Zwaan\thanks{\emph{Present address:}
   	Astrophysics Group,
   	School of Physics,
   	University of Melbourne,
   	Victoria 3010,
   	Australia} \\
        Kapteyn Astronomical Institute, P.O. Box 800, 9700 AV
        Groningen, The Netherlands\\
        email: mzwaan@physics.unimelb.edu.au
	} 
\date{Accepted ...
      Received ...}

\pagerange{\pageref{firstpage}--\pageref{lastpage}}
\pubyear{0000}

\begin{document}

\maketitle

\label{firstpage}

 \begin{abstract}
 Five galaxy groups with properties similar to those of the Local Group
have been surveyed for \hi\ clouds with the Arecibo Telescope.  In total
300 pointings have been observed on grids of approximately $2.5\times
1.5$~Mpc centred on the groups.  The 4.5$\sigma$ detection limit on the
minimal detectable \hi\ masses is approximately $7\times 10^6~\msol$
($H_0=65\,\hubble$).   All detections could be attributed to optical galaxies;
no significant detections of \hi\ clouds have been
made.  This null
result leads to the conclusion that the total \hi\ mass of intragroup
clouds must be less than 10 per cent of the total \hi\ mass of galaxy
groups and less than 0.05 per cent of the dynamical mass.  The recent
hypothesis that Galactic high velocity clouds are Local Group satellite
galaxies is highly inconsistent with these observations. 
 \end{abstract}


\section{Introduction}
 Groups of galaxies have been the subject of many \hi\ studies.  Famous
\hi\ maps of, for example, the M81 group (van der Hulst 1979, Appleton,
Davies \& Stephenson 1981, Yun, Ho \& Lo 1994) show that these groups
often host tidal \hi\ features (see also e.g., van Driel et al.  1992,
Li \& Seaquist 1994, Rand 1994, and Haynes, Giovanelli \& Chincarini
1984 for a review).  The incidence of \hi\ features
was quantified by Haynes (1981) who found that
six galaxy groups from a sample of 15 show \hi\ appendages from at least
one member galaxy.  These data might suggest that \hi\ filaments and
intragroup clouds are commonplace throughout galaxy groups, but such a
conclusion could be misleading since \hi\ surveys are normally
concentrated on the inner parts of galaxy groups and often on
interacting pairs within the groups.  Systematic searches throughout the
volumes occupied by the groups are rare because the projected sizes of
groups on the sky are too large to be covered by aperture synthesis
instruments.  A notable exception is the survey described by Lo \&
Sargent (1979) who systematically searched for \hi\ clouds throughout
the volumes around the M81, CVnI and NGC~1023 groups.  No clouds were
detected to the \hi\ mass limit of $4\times 10^7~\msol$, although
several new dwarf galaxies were discovered. 

Lo \& Sargent (1979) discussed their null-result in the context of
Galactic High Velocity Clouds (HVCs).  They concluded that HVCs are
unlikely to be intergalactic gas clouds in the Local Group (LG) because
they should have detected an equivalent population in the external
groups.  Similar conclusions were reached by Giovanelli (1981) and
Wakker \& van Woerden (1997). 
However, the idea that the HVCs instead
of being small clouds close to the Milky Way galaxy, are clouds of
primordial composition at LG distances has recently seen a revival. 
Originally proposed by Oort (1966), subsequently discussed by Verschuur
(1969) and Hulsbosch (1975), the idea has been refined by Blitz et al. 
(1999, hereafter BSTHB), who added a dark matter component to the
clouds. 
The problem with earlier considerations of the extragalactic
origin of HVCs was their derived distance.  Hulsbosch (1975) concluded
on the basis of the virial theorem that typical distances would be
approximately 10~Mpc, which places the clouds outside the LG.  BSTHB
show that if the HVCs are built from the same material that galaxies are
made of (typical baryon content \fb\ of 10 per cent), their distances
would reduce to $\sim 1$~Mpc.  Moreover, they show that the clouds'
distribution on the sky resembles that of LG dwarfs, and their
kinematics as an ensemble can be well modelled if they are distributed
throughout the LG.  Braun \& Burton (1999, hereafter BB) define a
subsample of 65 compact HVCs and come to essentially the same
conclusions about their subsample. Note that the BSTHB and BB models 
do not apply to the HVCs associated with the Magellanic Stream, which
are likely the result of tidal interactions between the
Milky Way and the Magellanic Clouds (Putman et al. 1998)

 Placed at LG distances, the HVCs have \hi\ masses of ${\sim}3 \times
10^7~M_\odot$ and typical diameters of 30 kpc.  In this picture,
approximately 500 to 1000 clouds are distributed throughout the LG,
together adding approximately $4\times 10^{10}~\msol$ to the LG \hi\
mass.  Interestingly, this number of HVCs is in reasonable agreement
with the number of mini halos that are predicted by numerical
simulations of the hierarchical LG formation (Klypin et al.  1999; Moore
et al.  1999).  Corrected for incompleteness, the BB sample comprises
$\sim 100$ clouds.  Using the updated average distance of 650~kpc (Braun
\& Burton 2000), the total \hi\ mass in their clouds would be
approximately $10^{9}~\msol$. 

Charlton, Churchill \& Rigby (2000) show that the statistics of Mg{\sc
ii} and Lyman limit absorbers in the spectra of quasars is in conflict
with the hypothesis that the HVCs are intragroup material (see also
Zwaan 2000).  For analogous HVC populations to exist around intervening
galaxies as well as in the LG, the typical distances would need to be
less than 200 kpc from the LG barycenter.  A very similar conclusion has
been reached by Zwaan \& Briggs (2000) who show that existing \hi\
surveys impose strong constraints on the existence of extragalactic
HVCs.  If \hi\ clouds exist in other groups or around galaxies, with
properties similar to those proposed for the LG, several instances
should already have been detected in large \hi\ surveys such as the
Arecibo survey discussed by Zwaan et al.  (1997). 

In this paper we present additional evidence by means of a targeted
survey of five galaxy groups with the Arecibo telescope.  The selection
of the targets is discussed in \S~\ref{selecttargets.sec} and the data
acquisition and analysis is summarised in \S~\ref{dataanalasysis.sec}. 
In \S~\ref{detections.sec} the detections are presented, and in
\S~\ref{groupanalysis.sec} we discuss the implications on the existence
of intragroup \hi\ clouds.  To enable direct comparison between the
surveyed groups and the LG, we adopt distances to the groups based on
their redshift velocities and a Hubble constant of $H_0=65\,
h_{65}\,\hubble$. 

\section{Sample selection} \label{selecttargets.sec}
 In order to make the targeted survey for \hi\ clouds most efficient, we
have selected the galaxy groups according to the following criteria:
 1) The distance to the groups must be such that the expected \hi\ cloud
diameters match the projected extent of the Arecibo beam.  BSTHB
estimate that the cloud diameters are approximately 30 kpc.  The $3'$
beam of the Arecibo Telescope subtends $d_{\rm beam}=0.87 D$~kpc at
a distance $D$ Mpc.  The optimal group distance at which the clouds fill
the beam is therefore 30 Mpc.  We have selected groups with radial
velocities in the range $\sim 1800~\kms$ to $3000~\kms$. 
 2) The declination must be in the range 10$^\circ$ to 30$^\circ$ so
that the groups are accessible to the Arecibo Telescope and can be
tracked for at least one hour. 
 3) To enable comparison with the LG, the global properties of the
groups, such as the integral \hi\ mass, luminosity, and dynamical mass,
must be comparable to those of the LG. 

  The list of galaxy groups compiled by Garcia (1993) was found to be
the most successful in meeting the above listed criteria.  This
catalogue was compiled from the LEDA\footnote{We have made use of the
Lyon-Meudon Extragalactic Database (LEDA) supplied by the LEDA team at
the CRAL-Observatoire de Lyon (France).} galaxy sample, and is basically
a cross section of groups found via a percolation method (Huchra \&
Geller 1982) and groups identified via the hierarchical clustering
method (Materne 1978).  Table~\ref{groupprops.tab} gives the measured
and derived properties of the selected groups.  The groups are named
after their brightest member.  

 \begin{table*}
 \begin{minipage}{15.2cm}
 \caption[]{Properties of surveyed groups}
 \label{groupprops.tab}
 \begin{tabular}{lccccccccc}
 \hline
 \noalign{\smallskip}
{(1)} & 
{(2)} & 
{(3)} & 
{(4)} & 
{(5)} & 
{(6)} & 
{(7)} & 
{(8)} & 
{(9)} &
{(10)} \\
{Name} & 
{$\alpha_{2000}$} & 
{$\delta_{2000}$} &
{$N_{\rm m}$} &
{$\langle v \rangle$} &
{$\sigma_r$} &
{$\log L_B$} &
{$\log \mhi$} &
{$\log M_{\rm dyn}$} &
{$R_0$} \\
{$\phantom{\int_{A_A}}$} &
{(h m)} &
{(${^\circ}$ $'$)} &
{} &
{($\kms$)} &
{($\kms$)} &
{($h^{-2}_{65} L_\odot$)} & 
{($h^{-2}_{65} M_\odot$)} & 
{($h^{-1}_{65} M_\odot$)} &
{($h^{-1}_{65} \rm Mpc$)} \\
 \noalign{\smallskip}
NGC    5798       & 14 56 & 30 21 &  3 &  1773 &   40 &   9.9 &   9.5 &  12.0 &  0.9 \\
NGC    5962       & 15 35 & 15 26 &  5 &  1872 &  126 &  10.8 &  10.0 &  13.0 &  1.9 \\
NGC    5970       & 15 37 & 12 14 &  4 &  1887 &   58 &  10.7 &  10.1 &  12.3 &  1.2 \\
NGC    6278       & 17 00 & 23 02 &  3 &  2911 &  104 &  10.5 &   9.6 &  12.8 &  1.7 \\
NGC    6500       & 17 53 & 18 16 &  6 &  2994 &  105 &  11.0 &  10.2 &  13.3 &  2.3 \\
NGC    6574       & 18 12 & 13 37 &  4 &  2279 &   28 &  10.9 &  10.2 &  12.2 &  1.1 \\      
 \noalign{\smallskip}
Local Group       & & & 3$^{\mathrm{a}}\!\!\!$ & & 61$^{\mathrm{b}}\!\!\!$ & 10.6$^{\mathrm{c}}\!\!\!$ & 10.1$^{\mathrm{d}}\!\!\!$ & 12.4$^{\mathrm{b}}\!\!\!$ & 1.2$^{\mathrm{b}}\!\!\!$ \\
 \noalign{\smallskip}
 \hline
 \end{tabular}

 {\medskip}
 (1) Most luminous member, 
 (2) and (3) Unweighted average RA and declination, 
 (4) Number of known member with measured redshifts,
 (5) Unweighted average radial velocity,
 (6) Radial velocity dispersion. The uncertainty on this is approximately
	$\sigma_r/ \sqrt{N_{\rm m}}$, 
 (7) Integral $B$-band luminosity,
 (8) Integral \hi\ mass,
 (9) Rough estimate of dynamical projected mass (see text),
 (10) Estimate of the `zero-velocity radius' (see text). \\
 $^{\mathrm{a}}$~Number of LG members included in the Garcia (1993)
         catalogue if the LG were to be at a distance of 30 Mpc, 
 $^{\mathrm{b}}$~Taken from Courteau \& van den Bergh (1999),
 $^{\mathrm{c}}$~Taken from van den Bergh (2000),
 $^{\mathrm{d}}$~Dwarfs from Mateo (1998) and giants from van den Bergh (2000).
\end{minipage}
\end{table*}

We have made crude estimates of the groups' dynamical masses by applying
the `projected mass estimator' which is discussed by Heisler, Tremaine
\& Bahcall (1985) and is defined as
 \begin{equation}
 M_{\rm PM} = \frac{32/\pi}{G(N_{\rm m}-1.5) \Sigma_i V_i^2 R_i},
 \end{equation}
 where $N_{\rm m}$ is the number of members, $V_i$ is the radial
velocity of member $i$ with respect to the group mean velocity, and
$R_i$ is the projected distance of member $i$ from the centre of the
group.  Since the groups have only 3 to 6 identified members, the errors
on the mass estimates are large (approximately $M_{\rm PM}/\sqrt{N_{\rm
m}}$).  The radius of the zero-velocity surface, $R_0$, beyond which
galaxies participate in the Hubble expansion, can be calculated via
(Sandage 1986)
 \begin{equation}
 R_0=\left(\frac{8 G T^2}{\pi^2}M_{\rm dyn}\right)^{1/3},
 \end{equation}
 where $T$ is the age of the group.  We take the ages of all groups to
be 14~Gyr. 

For comparison, we also give in Table~\ref{groupprops.tab} the most
recent published properties of the LG, taken from Mateo (1998), Courteau
\& van den Bergh (1999), and van den Bergh (2000).  For the LG, the
dynamical mass is estimated using the virial theorem, but it has been
shown by Heisler et al.  (1985) that the virial theorem and the
`projected mass estimator' give very similar results.  It is evident
from the table that the selected groups have properties not very
dissimilar from those of the LG: the selected groups are on average
equally gas rich, luminous, and massive and have a similar radial extent
as the LG.  Note that the centres of the NGC~5962 and NGC~5970 groups
are only $1.6 h^{-1}_{65}\rm Mpc$ apart and that their formal zero
velocity radii are overlapping.  Their difference in $v_{\rm hel}$ is
only 15~\kms.  It is suggestive that these two groups actually form one
gravitationally bound system.  If we assume that they form one group,
its dynamical mass would become $\log M_{\rm dyn}= 13.2 -\log h_{65}$,
and its zero velocity radius would be $2.3\, h^{-1}_{65} \rm Mpc$.  In
the remainder of this paper we regard the NGC~5962/5970 group as one
group.

\section{Observational strategy} \label{dataanalasysis.sec}
 Observations were carried out with the refurbished Arecibo\footnote{The
Arecibo Observatory is part of the National Astronomy and Ionosphere
Center, which is operated by Cornell University under a cooperative
agreement with the National Science Foundation.} Telescope during five
nights in the period from April 18 until 24, 1999 and on June 8 and 9,
1999.  The data were taken with the L-narrow Gregorian receiver which
has a measured system temperature of 32~K, and a gain of $10\rm K\,
Jy^{-1}$.  Spectra of 2048 channels were recorded for two polarisations
over a bandwidth of 12.5 MHz, centred on the frequency of the 21cm line
redshifted to the mean velocity of each galaxy group.  This setting
results in a total velocity coverage of $\sim 2600~\kms$ and a velocity
resolution of 1.3~\kms\ before Hanning smoothing.  Spectra were dumped
every 60 seconds. 

The pointings were arranged in rectangular grids centred on the galaxy
groups.  The grids were built from several series of five pointings on
lines of constant declination, where each pointing was separated by 5
minutes in hour angle.  The integration time per pointing was 4 minutes
and integrations were separated exactly 5 minutes in time.  The
remaining minute was used to slew the telescope back.  This strategy
insures that the same path on the telescope dish was traced during each
of the five pointing in a series.  No separate OFF scans were taken, but
for each scan a synthetic OFF scan was computed by averaging the other
four scans in the same series.  This synthetic OFF scan $y_j$ can be
written as
 \begin{equation}
 y_j= \frac{N \bar{x} -x_j }{N-1},
 \end{equation}
 where $N$ is the number of pointings in an array, $x_j$ is an
individual scan and $\bar{x}$ is the average of all scans in one series. 

 For this type of observation, this observing technique is superior in
its efficiency compared to a traditional observing scheme where separate
ON/OFF pairs are taken for each pointing.  For this ON/OFF scheme the
total noise depends on integration time as $\sigma \propto
\sqrt{1/{T_{\rm ON}}+1/{T_{\rm OFF}}}$.  Normally, the integration times
for the ON and the OFF scans are taken to be equal so that $\sigma
\propto 2/\sqrt{T_{\rm tot}}$, where $T_{\rm tot}=T_{\rm ON}+T_{\rm
OFF}$ is the total integration time needed for each pointing.  For our
strategy, where we make composite OFF scans from other scans in the
array, the noise $\sigma' \propto \sqrt{1/T'_{\rm tot} + 1/([N-1]T'_{\rm
tot})}$, where $N$ is the number of pointings in one series and $T'_{\rm
tot}$ is the time spend on each pointing.  From this we can derive
 \begin{equation}
 \frac{T_{\rm tot}}{T'_{\rm tot}} = \frac{4(N-1)}{N}\frac{\sigma'^2}{\sigma^2}.
 \end{equation}
 This shows that in our case, where $N=5$, the technique is a factor of
3.2 faster than traditional ON/OFF procedures.  In other words, in the
same amount of time a factor 1.8 higher signal to noise can be achieved. 

Fig.~1 shows a schematic view of the pointings that
were observed, together with the positions of the group galaxies.  The
circles indicate the sizes of the groups characterised by the
zero-gravity radius $R_0$.  The grids of pointings sparsely sample
rectangular areas of approximately $2.5 \times 1.5$ Mpc. 

 \begin{figure*} \epsfxsize=16cm
 \epsfbox[18 400 600 700]{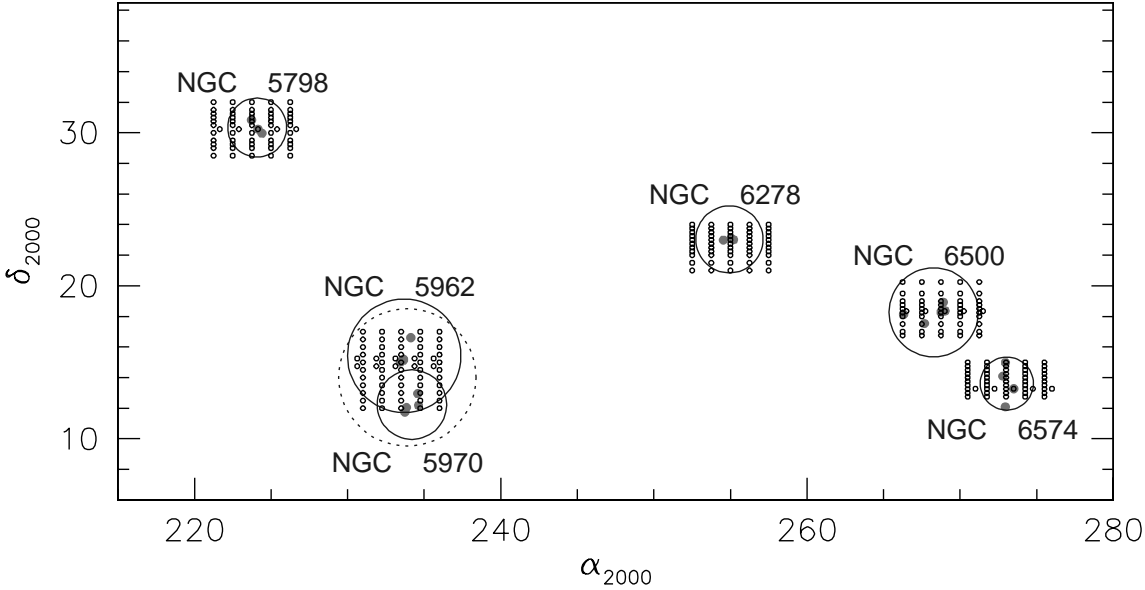}
\caption{Positions of groups and pointings on the sky.  The grey dots
represent the individual galaxies in the groups, the big circles show  
the surfaces of zero-gravity.  These are typically 1--2~Mpc.  For the
NGC~5962 and NGC~5970 groups their joint surface of zero-gravity is
also
indicated.  The groups are named after their brightest members.}
\label{pointings.fig} \end{figure*}

 Calibration of the data was performed by observing continuum sources
with known flux densities.  Separate scans and polarisations were
averaged and Hanning smoothed.  Polynomials were fitted to regions free
from obvious signals, and subtracted from the spectra in order to obtain
flat baselines.  A first order polynomial (linear baseline) was
generally found to be sufficient.  Each spectrum was then Gaussian
smoothed to resolutions 5, 10, 25 and 50~\kms.  The resulting noise
level was on average 0.75 mJy at 10~\kms\ resolution, which corresponds
to a minimal detectable \hi\ column density level of $1.2 \times
10^{18}~\icmsq$ ($4.5\sigma$).  The lowest detectable \hi\ mass at 30
Mpc was $7.1\times 10^6~\msol$ for a 10 \kms\ broad signal.  The \hi\
mass limit increases to $1.6\times 10^7~\msol$ for signals 50~\kms\
wide. 
 
All spectra were checked for $4.5\sigma$ peaks in the full resolution
and the smoothed spectra.  The list of 4.5$\sigma$ peaks was first
checked for false positives due to RFI by re-analysing the separate 60
second scans for both polarisations.  All peaks that could not
unambiguously be attributed to RFI were re-observed for confirmation on 8
and 9 June 1999.  The new observations were conducted following an
ON/OFF fashion, spending 8 minutes ON and 8 minutes OFF on each
position.  The resulting noise level for these observations was
therefore a factor $0.89$ lower than for the original observations
carried out in April 1999. 

\section{The detections} \label{detections.sec}
 Three pointings were deliberately aimed at the positions of known
galaxies.  This was done to check the positional accuracy of the
pointing method, and to obtain a confirmation of the flux calibration. 
The left three panels of Fig.~2 show the spectra of
NGC~5789, UGC~11168 and NGC~6500.  The latter one shows a secondary peak
at $\sim 3500~\kms$.  This turns out to be part of the \hi\ filament in
the galaxy pair NGC~6500/6501, that has been mapped with the WSRT by van
Driel, Davies \& Appleton (1988).  The authors describe the \hi\
structure as a `classical bridge and tail configuration of a double
galaxy interaction'.  This is obviously of tidal origin, and not a
primordial \hi\ cloud. 

The right panels of Fig.~2 show the three other
confirmed detections.  The first one is due to UGC~11037.  This galaxy
was not intentionally pointed at, but one of the pointings happened to
fall very close ($\sim 1.3'$) to this galaxy.  There is a similar
explanation for the detection in the spectrum in the central right
panel.  This pointing was only $2.6'$ separated from the centre of
NGC~6574.  The redshift of the signal agrees very well with the measured
redshift of NGC~6574. 

Finally, the lower right spectrum shows a detection at $\rm
15^h39^m00^s$, $12^\circ30'00'' \,(\alpha,\delta)$ in the direction of
the NGC~5962/5970 group.  There is no catalogued galaxy at this
position, but there is an obvious optical galaxy visible on the DSS,
only $3'$ to the north.  The observed \hi\ redshift is consistent with
membership in the NGC~5962/5970 group.  From the DSS we estimate the
brightness of the galaxy to be $m_B=17$ (assuming $B-V=0.5$).  The
number density of galaxies brighter that $m_B=17$ is approximately 4.0
per square degree (Metcalfe et al.  1995).  The probability of
encountering a galaxy of such brightness within a $3'$ radius is
therefore approximately 3 per cent.  Furthermore, the velocity width of
the signal is $150\pm 15~\kms$, a factor of $\sim 7$ larger than the
typical width of the HVCs in the Wakker \& van Woerden (1991) catalogue,
and broader than any known HVC.  We therefore conclude that this \hi\
signal is very likely associated with the optical galaxy.  Additional
21cm observations on this field are required to confirm this.

 \begin{figure}
 \epsfxsize=9.2cm \epsfbox[70 195 440 700]{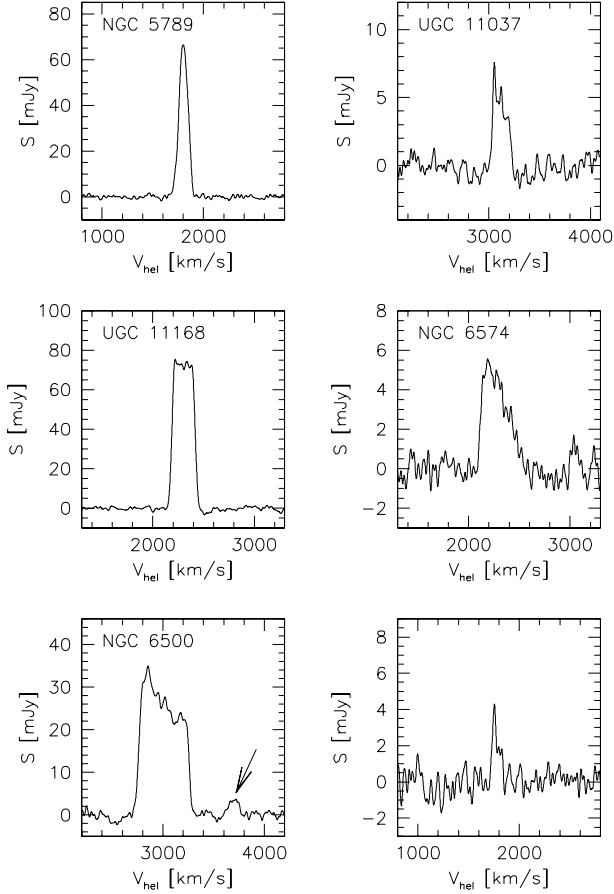}
 \caption{`Detections' in the Arecibo group survey. The left
three   
spectra are the result of pointings aimed at known galaxies. The
secondary peak in the spectrum of NGC 6500 is caused by a known
filamentary \hi\ structure in the NGC 6500/6501 pair. The upper right
and
middle right are serendipitous detections of known galaxies. The lower
right spectrum is probably the result of a nearby optical galaxy.
All spectra are smoothed to an effective resolution of 15~\kms.}
\label{spectrum.fig}
\end{figure} 

In summary, no \hi\ clouds of the type predicted by the BSTHB scenario
have been found.  Two detections were made that could not unambiguously
be identified as known optical galaxies.  One is a known tidal \hi\
filament in the NGC~6500/6501 pair, similar to the Magellanic Stream
seen in the Local Group (Putman et al.  1998).  The other detection is
very likely the result of a uncataloged member of the NGC~5970 group. 

\section{Space density of H\,{\sevensize\bf I} clouds} \label{groupanalysis.sec}
 We now use the null-result of the Arecibo group survey to derive upper
limits to the space density of \hi\ clouds in galaxy groups and discuss
the cosmological significance of intragroup \hi\ clouds. 

\subsection{HVCs as intragroup clouds}
 An explanation for Galactic HVCs that is of widespread current interest
is provided by BSTHB who suggest that most of the HVCs are actually
distributed throughout the LG and each cloud contains a few $\times
10^7~\msol$ of \hi.  We perform Monte Carlo simulations to test this
scenario by filling the five observed galaxy groups with synthetic
populations of clouds following a recipe outlined by BSTHB.  For the
cloud properties we use the measured solid angles $\Omega$, velocity
widths $\Delta V$, and average brightness temperatures $T_{\rm B}$ for
Galactic HVCs from Wakker \& van Woerden (1991).  Virial distances
$r_{\rm g}$ are calculated for each cloud individually.  The values of
$r_{\rm g}$ are directly proportional to the assumed ratio of baryon to
total mass \fb.  If \fb\ is 0.1, the virial distances $r_{\rm g}$ are
found to be approximately 1 Mpc.  At those distances, the distribution
of HVCs is in agreement with the kinematics of the LG, which was one of
the main motivations of BSTHB to propose the extragalactic HVC scenario. 

Within the groups, the clouds are placed at $r_{\rm g}$ from the group's
barycenter, in a random direction.  This situation would resemble that
in the LG, except that the substructure that BSTHB attribute to LG
dynamics is not simulated in the models of the external groups.  The
radial column density distribution for each cloud is first assumed to be
flat.  The average column density is calculated by taking $\langle
N_{\rm HI}\rangle =\mhi /(\pi r_{\rm cloud}^2)$, where \mhi\ is the
cloud \hi\ mass based on its value of $r_{\rm g}$ and its observed flux,
and $r_{\rm cloud}$ is the cloud radius, calculated from its measured
solid angle and $r_{\rm g}$.  For each model, a number $N$ clouds per
group is drawn randomly from the Galactic HVC parent population. 

The synthetic cloud ensembles are `observed' with patterns of beams
following the observational strategy of our survey.  A detection is
counted if the fraction of the flux of a cloud within the beam exceeds
the detection threshold.  The simulations are run with velocity
resolutions of  5, 10, 25 and 50~\kms, similar to the searching of the
real data.
 The same simulation is run 100 times for each group
in order to obtain reliable error estimates on the expected number of
detections. 
 
First, we assume that the number of clouds per group, $N$, is invariant
over the different groups.  If $N=100$, which is substantially lower
than the number of HVCs observed around the Milky Way Galaxy, the
expected total number of detections is $6 \pm 3$, where the error indicates
the $1\sigma$ variation around the mean.  Note that $N\approx 1000$ in
the BSTHB scenario and $N\approx 100$ in the BB scenario.  The expected
number of detections increases linearly with increasing $N$.  Next, we
drop the restriction that all groups contain an equal number of clouds
and instead scale $N$ with the dynamical mass of the group.  This seems
like a more logical thing to do since the \hi\ mass and luminosity are
also observed to scale in direct proportion to the dynamical mass. 
However, $N$ could of course be dependent on the dynamical state of the
groups.  In groups that have formed more recently, the primordial clouds
are likely to have been less efficient in merging than in older groups. 
We have no detailed information on the dynamical state of the groups,
and therefore simply assume that $N$ scales proportional to $M_{\rm
dyn}$.  We find that the expected total number of detections rises under this
assumption: $23\pm 8$ clouds are expected if the number of clouds in a
group with LG mass is $N=100$.  The reason for
this increase is that the average $M_{\rm dyn}$ for the external groups
is slightly higher than that for the LG. 

The conclusion from this exercise is that the hypothesis set forward by
BSTHB that HVCs are in-falling gas clouds in the LG is highly
inconsistent with the observations, unless the LG is an unusual group. 
If the LG is a representative group and the five surveyed external
groups are similar to the LG, our survey should have detected at least
30 clouds (assuming $N=500$).

\subsection{Constraints on intragroup H\,{\sevensize\bf I} cloud properties}
 A graphical representation of the constraints on intragroup \hi\ clouds
is presented in Fig.~3.  This figure shows the combined
constraints on the mean \hi\ mass of clouds, and the number of clouds in
each group.  The lines show 68, 90, 95, and 99 per cent confidence
levels at which the existence of a cloud population can be excluded. 
Again we have made use of the observed parameters of Galactic HVCs to
model cloud populations in the external groups and the number of clouds
$N$ is again scaled with $M_{\rm dyn}$.  For reference, the cloud
populations proposed by BSTHB and BB are indicated by hatched boxes, the
size of which reflects the uncertainty in the number and average \hi\
mass.  The horizontal arrow indicates the effect of changing the mean
distance of the BB clouds from 1~Mpc to 650~kpc from the Local Group
barycenter.  This latter value is preferred by Braun \& Burton (2000)
after they have estimated the distance to one HVC by comparing the
measured \hi\ column density and the angular size of the cool core. 
Both the BSTHB and the BB populations are inconsistent with the
observations at the $>99$ per cent confidence level. 

 \begin{figure}
 \epsfxsize=8.8cm \epsfbox[18 170 590 700]{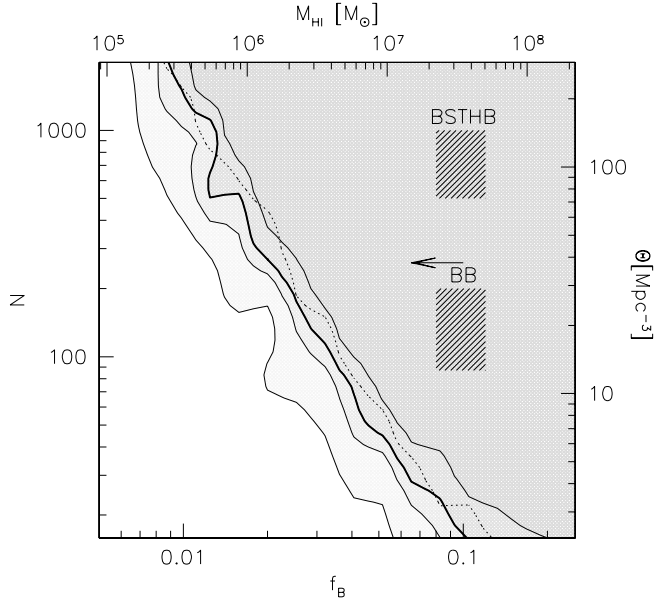}
 \caption{Combined constraint on the baryon fraction $f_{\rm B}$ and
the
number of clouds $N$ per group.  The number of clouds in each group
is normalised using the dynamical mass estimates, the number on the
vertical axis is the assumed number in the Local Group.  The average
\hi\ masses of the cloud populations are indicated on the top axis.  The
contours represent 68, 90, 95 (fat line), and 99 per cent confidence
levels on
the hypothesis that the existence of a group population can be rejected.
The dashed line is the 95 per cent confidence level assuming that the
clouds
can be described by a core-halo model in which 50 per cent of the flux
is in a
core with radius $R_{\rm cloud}/5$.}
\label{conf_dyn.fig}
\end{figure}
 
The distribution of \hi\ column densities in HVCs often show a core-halo
structure (Wakker \& van Woerden 1997).  Braun \& Burton (2000) present
high resolution WSRT imaging of six compact HVCs and show that the
morphology can be described by a diffuse halo that encompasses one or
more compact cores.  We test the influence of this morphology on the
detection efficiency in our survey by designing clouds with cores which
account for 50 per cent of the total flux and have a radius $R_{\rm
core}=R_{\rm cloud}/5$.  The remaining 50 per cent of the flux is
distributed over a halo with a flat \hi\ column density distribution. 
The 95 per cent confidence level on this population is indicated by a
dashed line.  It is clear that the detection efficiency is not
significantly changed by this modification of the cloud structure. 

Fig.~4 is similar to Fig.~3, but here
the number of clouds per group is non-variant.  Again in this case,
neither proposed population of clouds can be reconciled with our
observations. 

 \begin{figure}
 \epsfxsize=8.8cm \epsfbox[18 170 590 700]{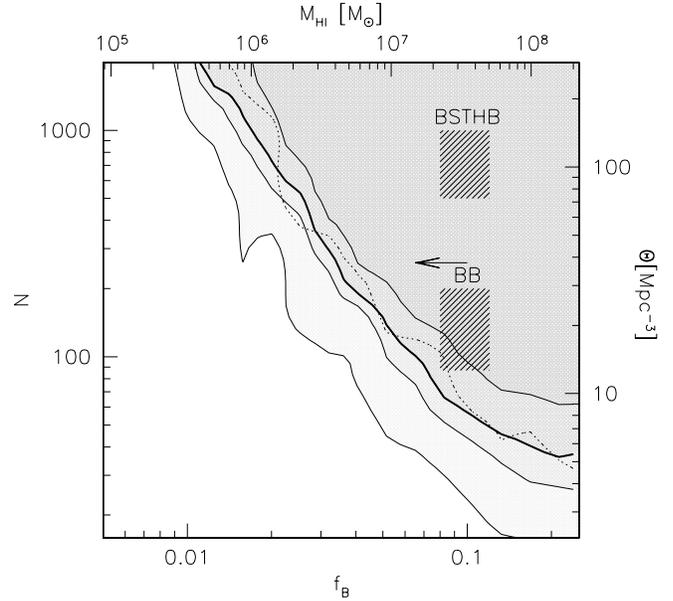}
 \caption{Same as Fig.~3, but here the number of
clouds per group, $N$, is equal for all galaxy groups.}
\label{conf_ns.fig}
\end{figure}

\subsection{Significance of intragroup clouds}
 How do these upper limits compare to the hierarchical formation
scenarios of galaxy groups? Klypin et al.  (1999) and Moore et al. 
(1999) show that in numerical simulations of a hierarchical universe the
relative amount of dark matter substructure halos is scale-invariant. 
The predicted relative number of dark matter halos is similar in
clusters, groups and galaxies.  However, only in clusters does the
predicted number of clumps agree with observed luminosity functions; on
galaxy and group scale the simulations predict an excess over the
observed number of satellites by a factor of 10, especially for halos
with circular velocities $<50~\kms$. 

One of the proposed solutions for this problem of missing satellites is
provided by the BSTHB hypothesis.  However, the evidence presented in
this paper, by Charlton et al.  (2000) and by Zwaan \& Briggs (2000)
seem to rule out this solution.  Only a very limited number of clouds
with $\mhi \sim 10^7~\msol$ could exist in galaxy groups.  A similar
conclusion has been reached by Verheijen et al.  (2000), who
systematically survey a region of the Ursa Major cluster of galaxies and
find no \hi\ clouds to a limit of $10^7~\msol$. 

From Fig.~3 we conclude that the intragroup \hi\ clouds
contribute a maximum of $1.0\times 10^9~\msol$ of \hi\ to the total
group mass.  This implies that no more than 10 per cent of all the \hi\
in groups can reside in clouds with masses greater than $\mhi=7\times
10^6~\msol$.  The \hi\ mass in clouds must be less than 0.05 per cent of
the total dynamical mass of the groups.  In the non-variant $N$ case,
these numbers rise to 20 and 0.1 per cent.  The dynamical mass of a
cloud population that is still permitted by the observations is more
difficult to constrain.  If we assume that the cold gas (\hi\ and \hei)
is the only baryonic component in the clouds, and the baryon fraction
$\fb = 0.10$ (a factor normally observed in galaxies and clusters,
Fukugita, Hogan \& Peebles 1998), then the total contribution of the
clouds to the dynamical mass of the groups must be less than 1 per cent. 
Note that Klypin et al.  (1999) estimate that the total mass in the
predicted dark matter satellites amounts to approximately 5 per cent of
the total group mass.  Such a high prediction can only be brought into
agreement with our survey results if the clouds' \fb\
is lowered to 0.02.  However, the median distance of the clouds
from the groups' barycentres would then reduce to $\sim 200$~kpc.  It is
not clear whether this is still consistent with the hierarchical model
predictions in which the dark matter satellites are distributed
throughout the groups. 

A solution to the problem of missing satellites might be that the cold
neutral gas is only a minor contributor to the total baryonic content of
the clouds making the \hi\ so insignificant that it can not be detected
in 21cm surveys.  This situation could occur if a large fraction of the
gas reservoirs in the satellites are ionised by the intergalactic
background.  Klypin et al.  (1999) and Moore et al.  (1999) discuss gas
ejection by early generation supernova-driven winds and inhibiting gas
cooling and star formation by photo-evaporation as possible explanation
of the absence of cold gas and stars in the satellites. 

Solutions of a different kind can be found by changing the predicted
number of clouds instead of modifying the baryons within the clouds. 
This can be achieved by suppressing of the primordial density
fluctuation spectrum on small scales (Kamionkowski \& Liddle 1999). 
Self-interacting dark matter (e.g., Spergel \& Steinhardt 1999) and
other dark-matter flavours (fluid dark matter, repulsive dark matter)
have been suggested as possible explanations for a less efficient
formation of small mass halos. 

\section{Summary}
 The conclusion reached by Lo \& Sargent (1979) that Galactic HVCs are
unlikely to be intergalactic gas in the Local Group (LG) remains sound
and intact under scrutiny of a new 21cm survey with the refurbished
Arecibo Telescope.  This new survey consists of 300 pointings in five
nearby galaxy groups and is sensitive to \hi\ masses of approximately
$7\times 10^6~\msol$, depending on the velocity spread and distance of
the signals.  Two detections have been made that are not unambiguously
caused by known optically selected galaxies.  One is a known tidal \hi\
filament in the NGC~6500/6501 pair, comparable to the Magellanic Stream
(Putman et al.  1998).  The other detection very likely originates in an
uncataloged member of the NGC~5970 group.  We therefore conclude that we
have made no significant detections of \hi\ clouds in galaxy groups. 

 We use this null-result to estimate constraints on the proposed
population of \hi\ clouds in groups, suggested by Blitz et al.  (1999)
and Braun \& Burton (1999).  These authors present a scenario in which
the Galactic high velocity clouds (HVCs) are actually distributed
throughout the LG at typical distances of a few hundred kpc to 1.5 Mpc. 
Fig.~3 shows the combined upper limits on the number of
clouds per galaxy group, and the average \hi\ mass on such clouds.  The
Blitz et al.  (1999) scenario can be ruled out with $>99$ per cent
confidence levels, assuming that the LG is typical of the five groups
studied here.  The integral amount of \hi\ in intragroup clouds is
typically less than 10 per cent of the groups' total \hi\ mass, and less
than 0.05 per cent of the total dynamical mass of the groups. 

The absence of clouds in groups seems to present a problem for
hierarchical structure formation scenarios that predict many satellites
within groups.  At present it remains unclear whether the solution to
this problem lies in modifying the descriptions of hierarchical
formation so that the predicted number of satellites drops, or that the
baryons in the clouds are simply hiding from detection.

\section*{acknowledgements}
 I thank F.  Briggs and the referee, V.  Kilborn, for useful comments
and discussion and K.  O'Neil for doing part of the observing for this
project.

\label{lastpage}

\end{document}